\documentclass[twocolumn,showpacs,preprintnumbers,amsmath,amssymb]{revtex4}

\usepackage{graphicx}
\usepackage{dcolumn}
\usepackage{bm}
\usepackage[latin1]{inputenc}

\begin{document}


\title{Enhanced Shot Noise in Tunneling through a Stack of Coupled Quantum Dots}

\author{P. Barthold$^1$}
\email{Barthold@nano.uni-hannover.de}
\author{F. Hohls$^{1,2}$}
\author{N. Maire$^1$}
\author{K. Pierz$^3$}
\author{R.J. Haug$^1$}
\affiliation{$^1$Institut f{\"u}r Festk{\"o}rperphysik,
Universit{\"a}t
Hannover, Appelstr. 2, D-30167 Hannover, Germany\\
$^2$Cavendish Laboratory, University of Cambridge, Madingley Road,
Cambridge CB3 0HE, Great Britain\\
$^3$Physikalisch-Technische Bundesanstalt, Bundesallee 100,
D-38116 Braunschweig, Germany}

\date{\today}

\begin{abstract}
We have investigated the noise properties of the tunneling current
through vertically coupled self-assembled InAs quantum dots. We
observe super-Poissonian shot noise at low temperatures. For
increased temperature this effect is suppressed. The
super-Poissonian noise is explained by capacitive coupling between
different stacks of quantum dots.

\end{abstract}



\maketitle
\begin{figure}
\includegraphics{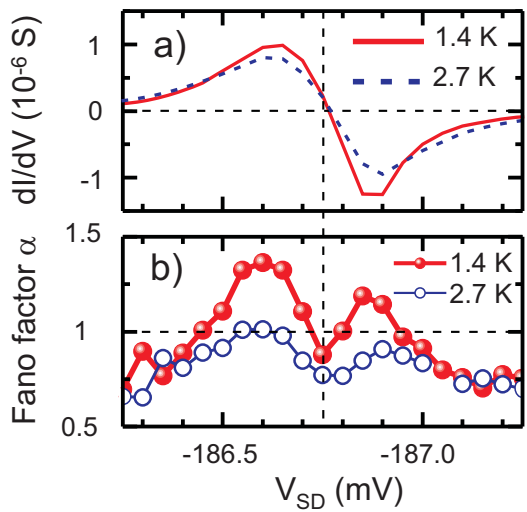}
\caption{\label{fig:diff-fano-temp} a) Differential conductance
for a peak in the I-V characteristic at T~$\approx$~1.4~K and
T~$\approx$~2.7~K. b)~The filled symbols show the Fano Factor
$\alpha$ at T~$\approx$~1.4~K and the open circles show $\alpha$
at T~$\approx$~2.7~K.}
\end{figure}

 Shot noise was introduced by Walter Schottky in 1918
while looking at the current fluctuations of vacuum
tubes~\cite{Schottky}. Due to the discreteness of the electrons
and the stochastic emission the current through the device
fluctuates around its average value. The corresponding shot noise
power is frequency independent and therefore called white noise. A
comparable effect is a single tunneling barrier in a semiconductor
device. The observed noise power density $S$ follows the same
expression $S=2eI$ where $I$ is the average current and $e$ is the
electron charge~\cite{MotBlanter}. It was found that the shot
noise power was reduced for tunneling through a double-barrier
structure $S<2eI$. Observations of a reduced noise power density
have been made for quantum well structures where the electrons
tunnel through a two-dimensional subband~\cite{LiFano,LiuFano} and
for systems with zero-dimensional
states~\cite{BirkFano,NauenFano01,NauenFano02,NauenTemp}. This
reduction is attributed to a negative correlation between the
tunneling events due to the finite dwell time of the resonant
state~\cite{ChenStrom,DaviesFano,KiesslichFano02}. A positive
correlation between the individual tunneling events has been
observed as well, which leads to an enhanced or so-called
super-Poissonian noise power~\cite{SuperIannaccone,SuperKuznetsov}
. The reasons for super-Poissonian shot noise depend on the
details of device structure. In Ref.~\onlinecite{SuperIannaccone}
enhanced shot noise is observed in a quantum well and explained by
Coulomb interaction and the shape of the density of states in the
well. In Ref.~\onlinecite{SuperKuznetsov} holes and the magnetic
field were taken into account to explain super-Poissonian shot
noise. It was shown that an impurity which is only weakly coupled
to the leads can modulate the tunneling current through a nearby
impurity leading to enhanced shot noise~\cite{Savchenko_Super}.
Super-Poissonian noise also has been studied experimentally
for triple-barrier resonant tunneling diodes~\cite{SuperPNoise}.\\
In this paper we study the shot noise properties of
resonant-tunneling through vertically coupled zero-dimensional
systems, so-called quantum dots (QDs). For given bias voltages the
current flows through a stack of vertical coupled InAs quantum
dots~\cite{Dipi_DD}. Therefore we deal with 3d-0d-0d-3d tunneling
in contrast to the above-mentioned experimental studies where the
tunneling takes place through a two-dimensional subband~\cite{SuperPNoise}.\\
\begin{figure}
\includegraphics[scale=0.6]{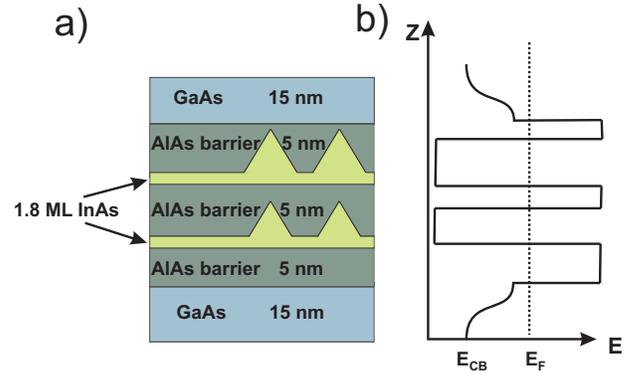}
\caption{\label{fig:sample} a) Growth scheme of the active part of
the device. b)~Conduction band structure $E_{CB}$ and the Fermi
energy~E$_{\mathrm{F}}$ of a stack of two QDs. If a positive bias
voltage is applied the electrons tunnel in positive z-direction
through the device.}
\end{figure}
The QDs were prepared by growing InAs on AlAs, the
lattice-mismatch of InAs and AlAs causes the formation of the
pyramidal QDs (Stranski-Krastanov-growth). The active region of
the sample consists of two layers of self-assembled InAs QDs (Fig.
\ref{fig:sample}, see also~\cite{Dipi_DD}). The bottom layer
contains QDs with a diameter of 10-15~nm and a height between 2
and 4~nm. The QDs that are formed in the second layer are aligned
to the dots in the first layer due to the remaining strain. The
QDs in the second layer are slightly larger than the QDs in the
first one~\cite{EiseleWachs}. These two layers of QDs are
surrounded by AlAs barriers. The three-dimensional emitter as well
as the collector are given by a 15~nm undoped GaAs spacer followed
by GaAs buffer with graded doping on both sides. The contacts are
realized by annealed Au/Ge/Ni/Au contacts. About one million QDs
are placed randomly on the area of an etched diode structure of
40~$\mu$m$~\times$~40~$\mu$m. It has been shown that in similar
samples with one layer of self-assembled QDs only a small fraction
of these QDs participates in the electronic
transport~\cite{QD-Hapke}. It seems reasonable that in stacked QDs
even less QDs allow electronic transport through the
diodes~\cite{Dipi_DD}.\\
To investigate the noise properties the sample is placed in a
$^{4}$He bath cryostat. During the measurements the sample is
always immersed in liquid helium. We can control the temperature
between 1.4~K and 4~K. The external bias voltage $V_{SD}$ is
applied between the source and the drain electrodes using a
filtered dc-voltage source. The detected signal is amplified by a
low-noise current amplifier with a 3~dB bandwidth of 10~kHz. The
signal is fed into a Fast Fourier-Transform analyzer (FFT) to
characterize the frequency spectra in a range from 16~Hz to
12.8~kHz with 16~Hz resolution. In addition we measure the
stationary current.\\
\begin{figure}
\includegraphics{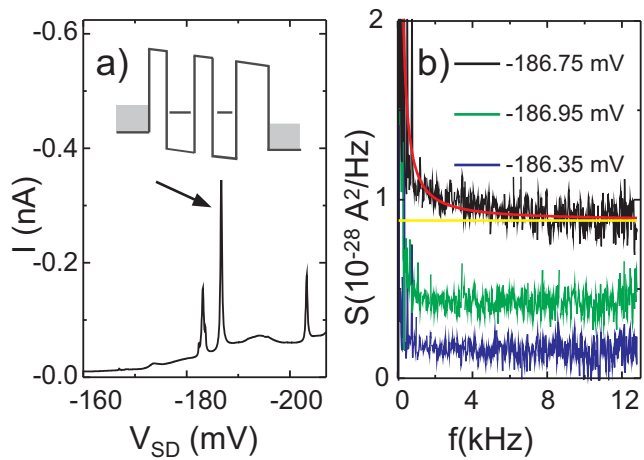}
\caption{\label{fig:ui-spectra} a) Current-voltage characteristics
of the sample at $T\approx$~1.4~K. The arrow marks the peak which
is discussed further. Inset: Schematic bandstructure of the
coupled quantum dots where resonant tunneling is observed.
b)~Typical noise spectra of the sample for different bias
voltages. The solid line demonstrates the results of fitting the
function $A/f+S_{0}$ to the spectrum at $V_{SD}=-186.75$~mV. The
horizontal line shows $S_{0}$.}
\end{figure}
Figure~\ref{fig:ui-spectra}a  demonstrates a part of the
current-voltage characteristic of the sample at $T\approx$~1.4~K .
The curve shows three well defined peaks. To understand this
behavior we take a look at the inset. It shows the schematic
bandstructure of the sample with external voltage $V_{SD}$
applied. On the left side we see the three-dimensional emitter
followed by three barriers. In between these barriers are two
quantized states, the ground states of the vertically coupled InAs
quantum dots. On the right side the three-dimensional collector is
shown. If there is no external voltage applied to the sample, the
two states of the QDs are above the Fermi energy of the emitter
and the collector. By applying an external voltage $V_{SD}$ we
lower the energy levels of the dots and the collector in
comparison to the Fermi energy of the emitter. At a certain
voltage the two states in the QDs are in resonance. At this point
electrons are able to tunnel from the emitter into the first and
then into the second QD and from there into unoccupied states of
the collector. This process of sequential tunneling causes a peak
in the current-voltage characteristic. If the states in the QDs
are not resonant an electron has to tunnel through the three
barriers at once creating a small leakage current. Such a behavior
is described in detail in Ref.~\onlinecite{Dipi_DD} and
\onlinecite{Bryllert_DD}.
\\In Fig.~\ref{fig:ui-spectra}b we present
typical noise spectra of the sample. Spectra as the ones for
$V_{SD}=-186.35$~mV and $V_{SD}=-186.95$~mV show a frequency
independent behavior above $4$~kHz. At very low frequencies the
spectra show typically a 1/f-behavior. We fit the function
$A/f+S_{0}$ to the spectra to obtain information about the
amplitude of the shot noise $S_{0}$. By comparing $S_0$ to the
averaged noise spectra between 5~kHz and 12.8~kHz we find that
$S_0$ provides reliable
information.\\
\begin{figure}
\includegraphics{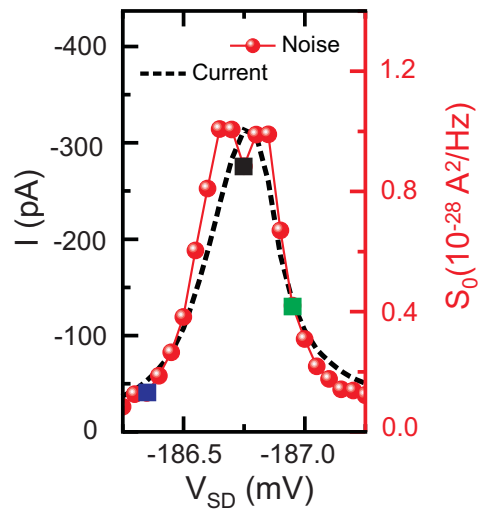}
\caption{\label{fig:i-v-noise} Blow up of the peak marked in
Fig.~\ref{fig:ui-spectra} in the current-voltage characteristic
measured at $T\approx$~1.4~K (dashed line, left axis) and shot
noise amplitude $S_0$ as given by fitting the function $A/f+S_{0}$
to the noise spectra shown in Fig.~\ref{fig:ui-spectra}~(dots,
right axis). The squares correspond to the spectra shown in
Fig.~\ref{fig:ui-spectra}b. The scale of the right axis was chosen
such that the dashed line corresponds on this axis to full shot
noise $S_{full}=2eI$ as expected for a single tunneling barrier
structure (Fano factor $\alpha = 1$).}
\end{figure}
In Fig.~\ref{fig:i-v-noise} we take a closer look at the marked
peak in the current-voltage characteristic
(Fig.~\ref{fig:ui-spectra}a). The left axis in
Fig.~\ref{fig:i-v-noise} demonstrates the current $I$ (dashed
line). The right axis shows the noise density $S_0$(line with
dots). The squares correspond to the spectra shown in
Fig.~\ref{fig:ui-spectra}. The scale on the axis is chosen such
that the dashed line also corresponds to full shot noise
$S_{full}=2eI$ on the right axis. Coming from low voltages
$V_{SD}$ the shot noise power $S$ is slightly suppressed. At
$V_{SD}\approx-186.4$~mV the shot noise power starts to be larger
than the full (Poissonian) shot noise (dashed line). We observe an
enhanced shot noise power at both sides of the peak while the shot
noise is reduced at the maximum of the current peak. Above
$V_{SD}\approx -186.95$~mV the shot noise begins to be suppressed
again.
\\In Fig.~\ref{fig:diff-fano-temp}a the differential
conductance of the sample is shown for $T\approx~1.4$~K and
$T\approx~2.7$~K for the same range of $V_{SD}$ as used in
Fig.~\ref{fig:i-v-noise}. To compare the differential conductance
with the shot noise amplitude we display the so-called Fano factor
$\alpha$ in the figure beneath (Fig.~\ref{fig:diff-fano-temp}b).
The Fano factor compares the measured shot noise power $S_0$ and
the full shot noise $2eI$: $\alpha := S_0/(2eI)$. The filled
symbols in Fig.~\ref{fig:diff-fano-temp}b show the Fano factor
$\alpha$ for
$T\approx 1.4$~K while the open symbols show $\alpha$ for $T\approx 2.7$~K.\\
For $T\approx1.4$~K the Fano factor $\alpha$ displays a distinct
double-peak behavior. The peaks in the Fano factor are clearly
reduced at $T\approx 2.7$~K, however, the double-peak structure is
evident for both temperatures. At lower voltages than
$V_{SD}\approx -186.45$~mV the Fano factor $\alpha$ is smaller
than 1 for both temperatures. While the differential conductance
increases the Fano factor rises to values above 1. The maximum in
the differential conductance coincides with the first maximum of
the Fano factor $\alpha$ at $V_{SD}\approx -186.65$~mV at both
temperatures where the values of the Fano factor are $\alpha
\approx 1.40$ at $T\approx 1.4$~K and $\alpha \approx 1.03$ at $T
= 2.7$~K. Together with the decreasing differential conductance
the Fano Factor drops to its minimum where the differential
conductance is zero. This minimum is nearly independent of the
temperature and has a value of about 0.89 for $T\approx 1.4$~K and
0.78 for $T\approx 2.7$~K. With further rise of the bias voltage
the differential conductance shows its minimum where the Fano
factor $\alpha$ reaches its second local maximum at $\alpha\approx
1.22$ at $T\approx 1.4$~K and $\alpha\approx 0.9$ at $T\approx
2.7$~K.
Comparing the temperature dependence of the Fano factor to the
differential conductance we find that the latter shows a much
smaller temperature dependence.\\
The observed double-peak structure of super-Poissonian noise is
quite astonishing, because theoretical approaches (like
Ref.~\onlinecite{Super-P-Kiesslich}) describing only a pair of
vertically coupled QDs expect a Fano factor $\alpha<1$. In
contrast, theoretical models that take coupling mechanisms to
other systems into account predict super-Poissonian shot noise.\\
A Monte Carlo and analytic simulation of four metallic dots
arranged as two parallel, only capacitively coupled double dots
showed an enhanced Fano factor with a double-peak structure very
similar to our experiment~\cite{enhancement_Gattobigio}. Here the
enhancement of the shot noise stems from the Coulomb interaction
of neighboring QDs. Each peak in the Fano factor occurs where the
difference between both currents through the parallel metallic QDs
reaches its maximum. Gattobigio et
al.~\cite{enhancement_Gattobigio} showed that a so-called locking
effect occurs due to electrostatic coupling between two pairs of
dots. A similar scenario is quite likely in our device as it
consists of a large number of stacked self-assembled dots. It is
reasonable that two or more stacks might conduct at the same bias
voltage.\\
Gattobigio et al.~\cite{enhancement_Gattobigio} have shown that
the value of the maxima in the Fano factor depends on the ratio
between the tunneling resistances of each parallel pair of QDs. A
ratio of 1 leads to a symmetric double-peak in the Fano factor,
other tunneling resistances lead to an asymmetric Fano factor.
This could be an explanation for the strong temperature dependence
of the Fano factor that is observed. It is possible that this
ratio is changed in our device by changing the temperature which
would influence the maximum and the ratio of the two peaks in the
Fano factor. With rising ratio of the tunneling resistances the
asymmetry of the double-peak in the Fano factor growths. Supposing
that the temperature dependence of the QD which contributes less
current is larger than the other, it would be easy to understand
why the current peak shows only slight changes with rising
temperature while the Fano factor depending on the ratio of the
two tunneling
resistances shows a strong temperature dependence.\\
Another coupling between two systems that causes an enhancement in
the shot noise was reported in Ref.~\onlinecite{Giant-Fano-Oppen}.
Koch et al.~\cite{Giant-Fano-Oppen} have shown that the Fano
factor can rise to giant values in transport through a single
molecule where a strong coupling between phonons and electrons
exists. This calculations were done in the Franck-Condon-blockade
regime. It is also possible that such a phonon-electron coupling
might play a role in
our device.\\
For a quantitative description of the observed shot noise further
detailed calculations for such a device are necessary.\\
In conclusion, we presented measurements of shot noise on
self-assembled coupled InAs QDs. To characterize the shot noise we
introduced the Fano factor~$\alpha$. The Fano factor~$\alpha$
shows a double-peak structure with maxima above 1 i.e.
super-Poissonian noise which is explained in terms of coupling
between different stacks of quantum dots.\\
We acknowledge financial support from BMBF.

\end{document}